\documentclass[a4paper]{jpconf}
\usepackage{graphicx}
\usepackage{amsmath,amssymb,amsfonts}
\begin{document}
\title{Monte Carlo simulation of entropy-driven pattern formation in two-dimensional system of rectangular particles}

\author{Yuri~Yu~Tarasevich, Andrei~V~Eserkepov, Valentina~V~Chirkova, Valeria~A~Goltseva}
\ead{tarasevich@asu.edu.ru}
\address{Astrakhan State University, Astrakhan, 414056, Russia}

\begin{abstract}
We simulated random walk of rectangular particles on a square lattice with periodic boundary conditions. Two kind of particles were investigated, viz., so-called ``blind'' and ``myopic'' particles. We found that steady state patterns occurred only for some values of the ratio $L_x/L_y$ where $L_x$ and $Ly$ are the linear sizes of the system and only for ``needles'', i.e., the particles of size $1 \times k$. Different patterns were observed for ``blind'' and ``myopic'' particles.
\end{abstract}

\section{Introduction}\label{sec:intro}
Recently, diffusion-driven pattern formation in a 2D discrete system has been studied by means of Monte Carlo (MC) simulations~\cite{Lebovka2017PRE,Tarasevich2017JSM,Patra2018PRE}. The observed behaviour resembled the pattern formations in thin layers composed of elongated granules subjected to vibrations~\cite{Boerzsoenyi2013SM,Muller2015PRE,Gonzalez2017SM}.

The present conference paper is devoted to a detailed analysis of how the different kinetics of particles (``blind'' and ``myopic'' particles) affects pattern formation in a two-species diffusion system. Moreover, we have examined the effect of shape of particles and shape of the substrate on the such pattern formation.

The rest of the paper is organised as follows. In section~\ref{sec:methods}, the technical details of the simulations are described and all necessary quantities are introduced. Section~\ref{sec:results} presents our principal findings. Section~\ref{sec:concl} summarises the main results.

\section{Methods}\label{sec:methods}
We used square lattices of size $L_x \times L_y$ with periodic boundary conditions.
We considered the rectangular particles of size $k_x \times k_y$ ($2 \times 2$, $4 \times 4$, $8 \times 8$, $2 \times 12$, $2 \times 16$). A special attention was paid to the case $1 \times k$ (``needles'' or linear $k$-mers). The initial state of the system under consideration corresponded to the jamming. The jammed state was generated using the random sequential adsorption~\cite{Evans1993RMP}. Even one additional particle cannot be added to the jammed system due to absence of the appropriate empty space.

Hard-core interaction between particles were assumed. Two kind of particle movement were examined. The first kind of the particles corresponds to ``blind'' particles, i.e., a particle chooses one of four possible directions and attempts to shift in this direction. If the attempt is unsuccessful, other particle attempts to move.
The second kind corresponds to ``myopic'' particles~\cite{Mitescu1983}, a particle is ``smart'', viz., it  chooses one of four possible directions and attempts to shift in this direction. If the attempt is unsuccessful, the particle  chooses other direction until the first successful attempt or when all four directions would examined.

In the present study, we utilize four main quantities, i.e., the normalized number of clusters, $n$, the degrees of freedom (DoF) per particle, $f$, the order parameter, $s$, and the relative number of the contacts, $n_{xy}^\ast$.

In percolation theory, a cluster is a group of neighbouring occupied sites~\cite{Stauffer}. In our particular case, a cluster is a group of particles of the same orientation connected with one another. Two particles are considered to be connected when there is at least one pair of neighbouring sites that belong to different particles. The clusters built of horizontal particles and clusters built of vertical particles were counted separately and then their numbers were averaged. We used the Hoshen--Kopelman algorithm~\cite{HK76} to count the clusters. The normalised number of clusters, $n$, is the current number of clusters divided by the value at the initial state, $t_\text{MC} = 0$. %

We define the DoF as number of possible unit movements of a particle from its current location.

We exploited the order parameter defined as
\begin{equation}\label{eq:s}
s = \frac{\left|N_y - N_x\right|}{N},
\end{equation}
where $N_x$ and $N_y$  are the numbers of sites belonging to the particles oriented along the $x$ and $y$ directions, respectively, and $N=N_y + N_x$ is the total number of the occupied sites. This order parameter, $s$, was calculated within a sliding window of $w \times w$ sites and then averaged over the entirely set of windows, i.e., over $L_x \times L_y$ windows except empty windows. We utilize $w=2^n$, where $n = 1,2,\dots, 7$. In all cases, we considered only isotropic systems, i.e., vertically and horizontally oriented particles  were equiprobable in their  deposition, hence, the order parameter calculated in the window of size $L_x \times L_y$ equals to zero, $s=0$.

The relative number of interspecific contacts $n_{xy}^{\ast} = n_{xy} / ( n_{xy} + n_x + n_y )$, where $n_{xy}$ is the number of interspecific contacts between the different sorts of particles (i.e., horizontal--vertical), $n_x$ and $n_y$  are the numbers of intraspecific contacts between particles of the same kind (i.e., horizontal--horizontal and vertical--vertical, respectively).

\section{Results}\label{sec:results}
We investigated the random walk of particles with the different aspect ratios $k_x/k_y$. Since the pattern formation was observed only for ``needles'', i.e., the particles of size $1 \times k$, the text below is devoted only to this particular particle shape.

\subsection{Comparison of ``blind'' and ``myopic'' kinetics}
This subsection is devoted to  ``needles'', i.e., the particles of size $1 \times k$, on the lattices of size $L_x=L_y=L$ with fixed ratio $L/k = 32$ when other is not explicitly indicated.

\begin{figure}[ht]
\begin{minipage}[t]{0.45\textwidth}
\includegraphics[width=\textwidth]{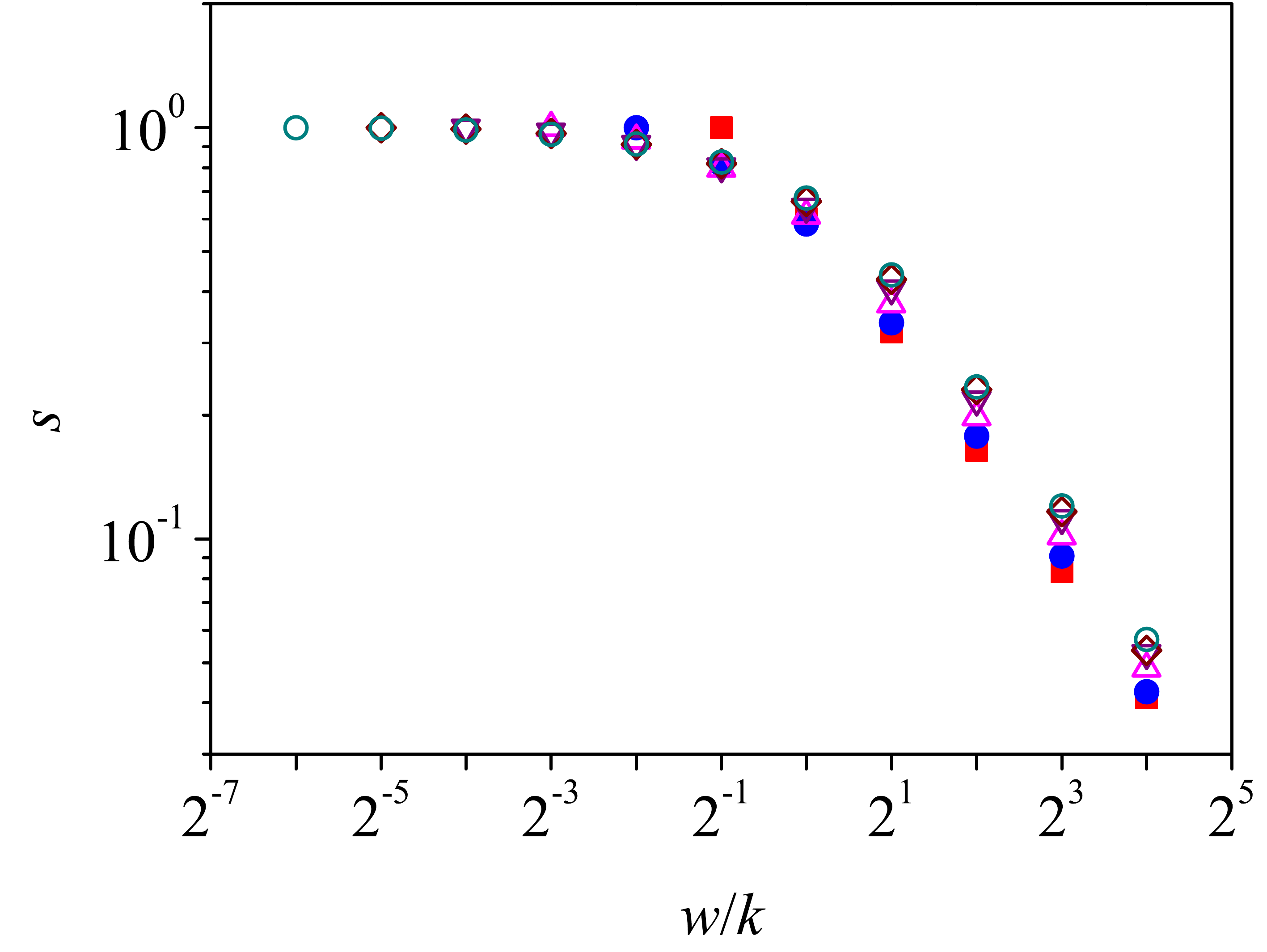}%
\caption{\label{fig:svswdivk}Examples of dependencies of the local order parameter, $s$, at the initial jammed state ($t_\text{MC} = 0$) vs the ratio $w/k$, where $w$ is the size of the window  in which the local order parameter was calculated, $k$ is the ``needle'' length.
$\fullsquare$ $k=2$,
$\fullcircle$ $k=4$,
$\opentriangle$ $k=8$,
$\opentriangledown$ $k = 16$,
$\opendiamond$ $k=32$,
$\opencircle$ $k=64$.
 }
\end{minipage}
\hfill
\begin{minipage}[t]{0.45\textwidth}
\includegraphics[width=\textwidth]{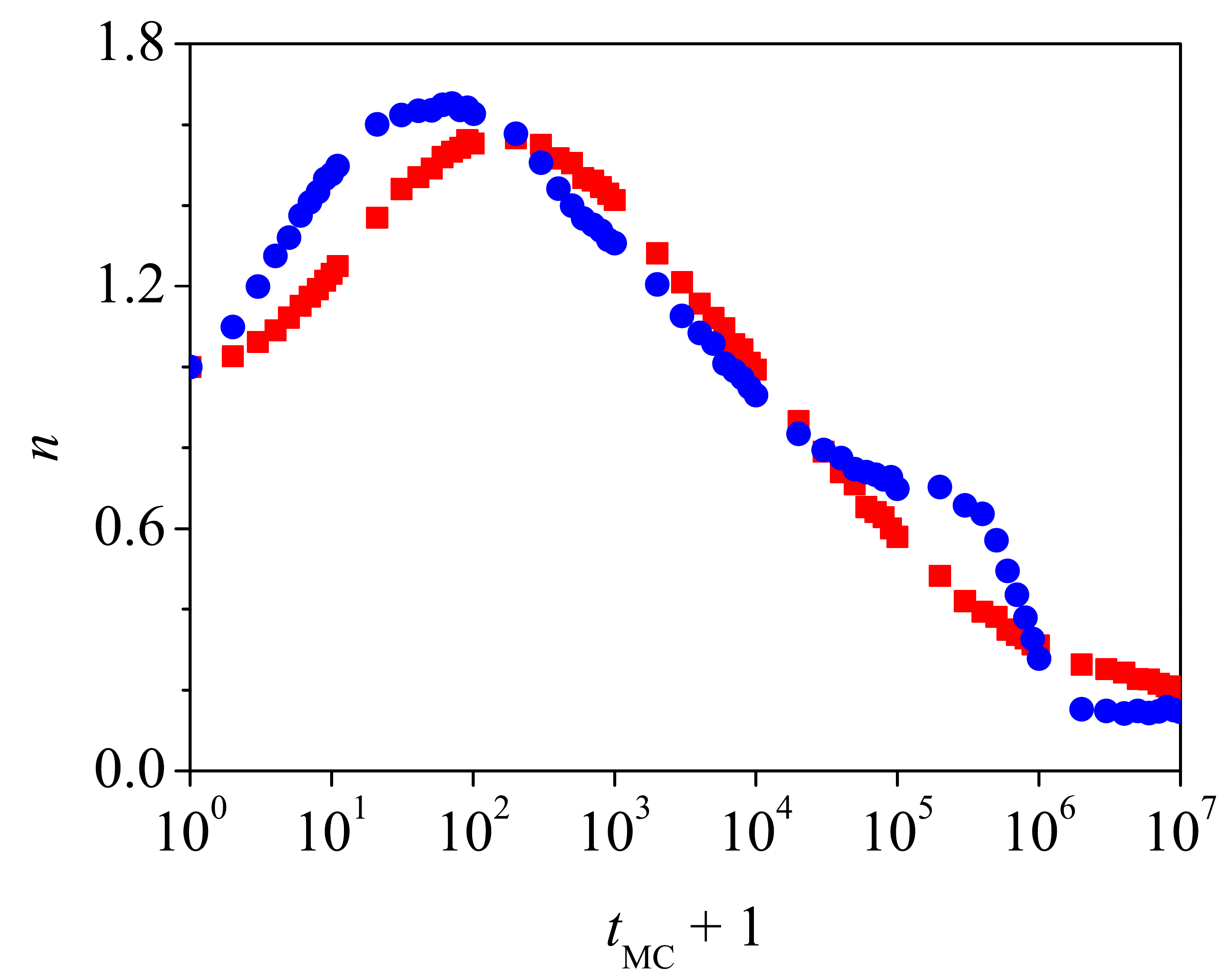}
\caption{\label{fig:SGridClustersNormL256k8}Examples of dependencies of the normalized
number of clusters, $n$, vs MC steps, $t_\text{MC}$.  $k = 8$. $\fullcircle$ ``myopic'' particles, $\fullsquare$ ``blind'' particles.}
\end{minipage}%
\end{figure}

\Fref{fig:svswdivk} demonstrates examples of the local order parameter, $s$, vs the ratio $w/k$ for some particular values of $k$ and $t_\text{MC} = 0$. Here, $w$ is the size of the window  in which the local order parameter was calculated. The order parameter evidenced that, in the initial jammed state, particles formed stacks of typical size $k \times k$. This stack structure was unstable, i.e., the random walk of needles destroyed this structure and brought the system to a new state. The transitions to a new state were different for ``blind'' and ``myopic'' particles. For both kinds of particles, the normalized number of clusters increased up to $t_\text{MC} \approx 10^2$ then decreased up to $t_\text{MC} \approx 10^6$ (``myopic'' particles) or $t_\text{MC} \approx 10^7$ (``blind'' particles) (\fref{fig:SGridClustersNormL256k8}). Nevertheless, for ``myopic'' particles, the maximum of the normalized number of clusters reached earlier. For this kind of particles, the curve $n\left(t_\text{MC}\right)$ had a step when $10^5 \lesssim t_\text{MC} \lesssim 10^6$.

\begin{figure}[!hb]
\begin{minipage}[t]{0.45\textwidth}
\includegraphics[width=\textwidth]{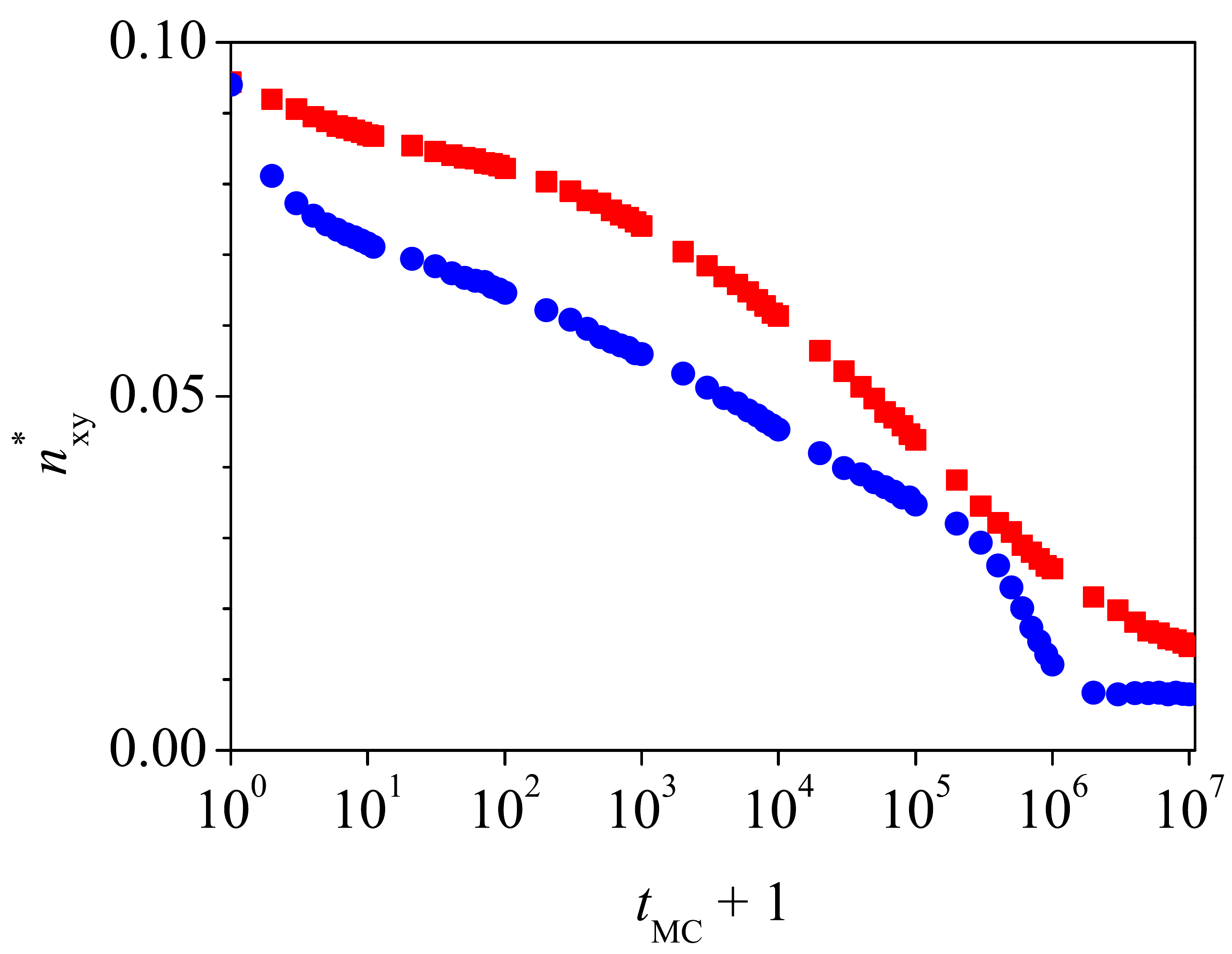}
\caption{\label{fig:SGridnxystarL256k8}Examples of dependencies of the relative number of the contacts, $n_{xy}^\ast$, vs MC steps, $t_\text{MC}$. $k= 8$. $\fullcircle$ ``myopic'' particles, $\fullsquare$ ``blind'' particles.}
\end{minipage}
\hfill
\begin{minipage}[t]{0.45\textwidth}
\includegraphics[width=\textwidth]{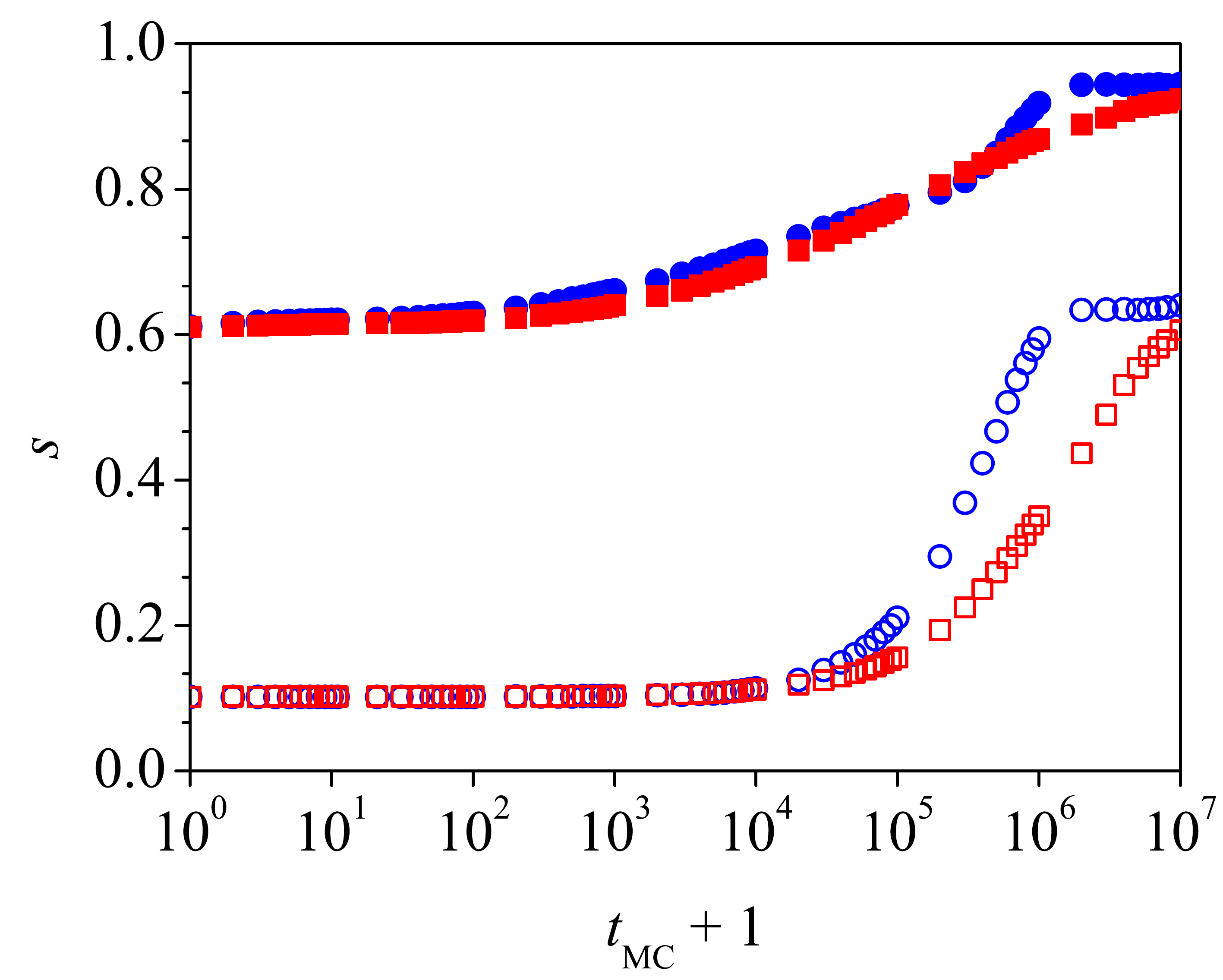}%
\hfill
\caption{\label{fig:SGsvstL256k8}Examples of dependencies of the local order parameter,
$s$, calculated in windows of size $8\times 8$ (closed symbols) and $64\times 64$ (open symbols) sites vs MC steps, $t_\text{MC}$. $k = 8$. $\fullcircle$, $\opencircle$ ``myopic'' particles, $\fullsquare$, $\opensquare$ ``blind'' particles.}
\end{minipage}
\end{figure}

\Fref{fig:SGridnxystarL256k8} demonstrates temporal dynamics of the relative number of the contacts, $n_{xy}^\ast$. For both kinds of particles, the system tended to a state with small number of contacts between particles of different orientations, i.e., a segregation of particles with different orientations occurred. This state was reached at $t_\text{MC} \approx 10^6$ in the case of ``myopic'' particles and at $t_\text{MC} \gtrsim 10^7$ in the case of ``blind'' particles. \Fref{fig:SGsvstL256k8} shows that the system evolved  from short-range order (ordered regions or stacks of typical size $k \times k$) to the long-range order (ordered regions of typical size $L/4$).

\begin{figure}[!htbp]
\begin{minipage}[c]{0.45\textwidth}
\includegraphics[width=\textwidth]{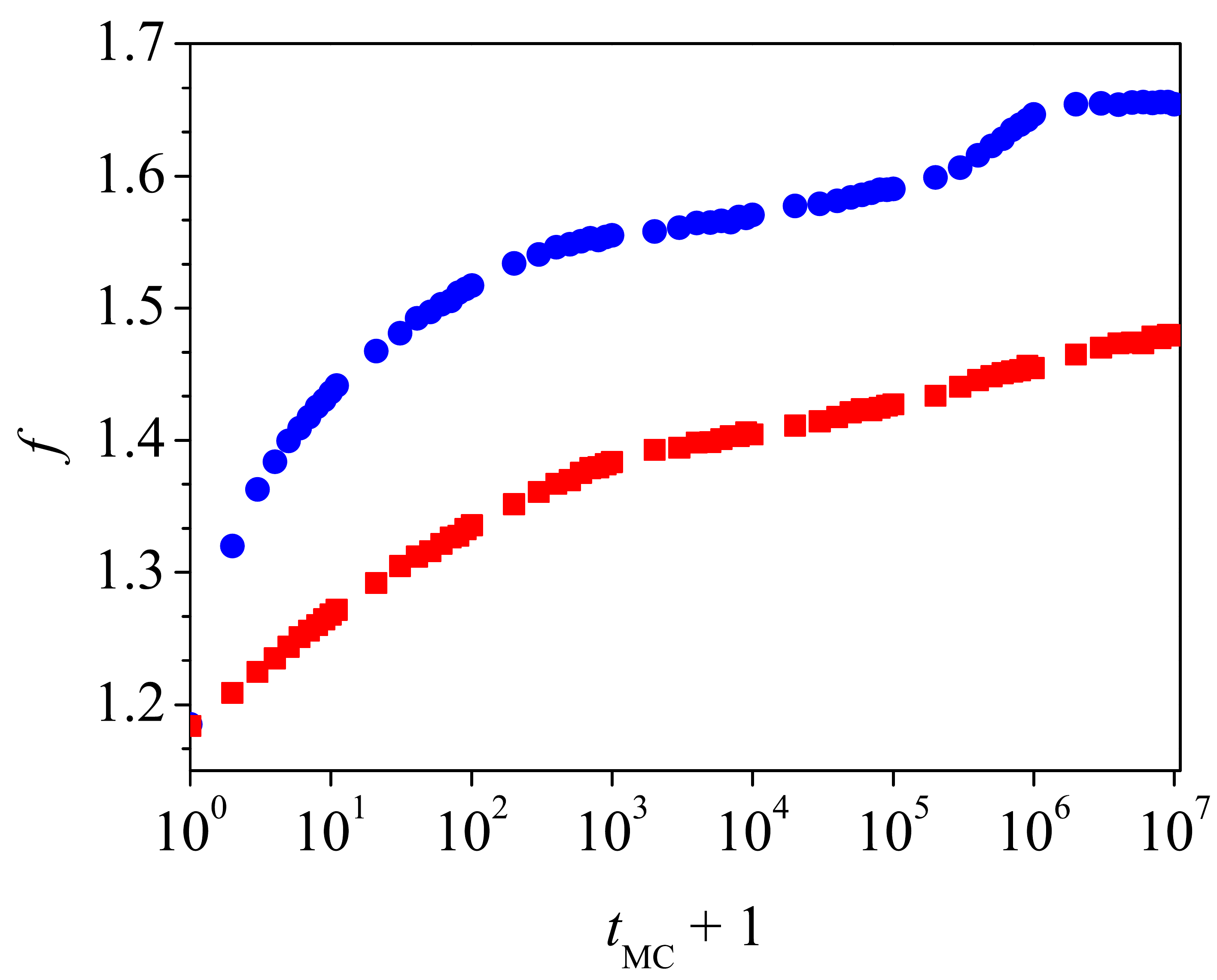}
\end{minipage}\hfill
\begin{minipage}[c]{0.45\textwidth}
\caption{\label{fig:SGfreek8}Examples of dependencies of the degrees of freedom, $f$, vs MC steps, $t_\text{MC}$. $k = 8$. $\fullcircle$ ``myopic'' particles, $\fullsquare$ ``blind'' particles.}
\end{minipage}
\end{figure}

For both kinds of particles, the random walk led to increase of the freedom of particles (\fref{fig:SGfreek8}). Due to hard-core interaction between particles, pattern formation should be treated as entropy-driven. Formation of the macroscopic order was accompanying with increase of the microscopic disorder, i.e., in the initial jammed state, a particle had less possibilities to change its location in compare with any succeeding state.

\subsection{Effect of the substrate shape}
A special attention has been paid to the effect of the lattice shape on the pattern formation in the systems of the ``myopic'' particles. We examined the lattices with different ratio $L_x/L_y$.

\begin{figure}[!htbp]
\begin{minipage}[b]{0.6\textwidth}
\centering
($a$)\includegraphics[width=0.425\textwidth]{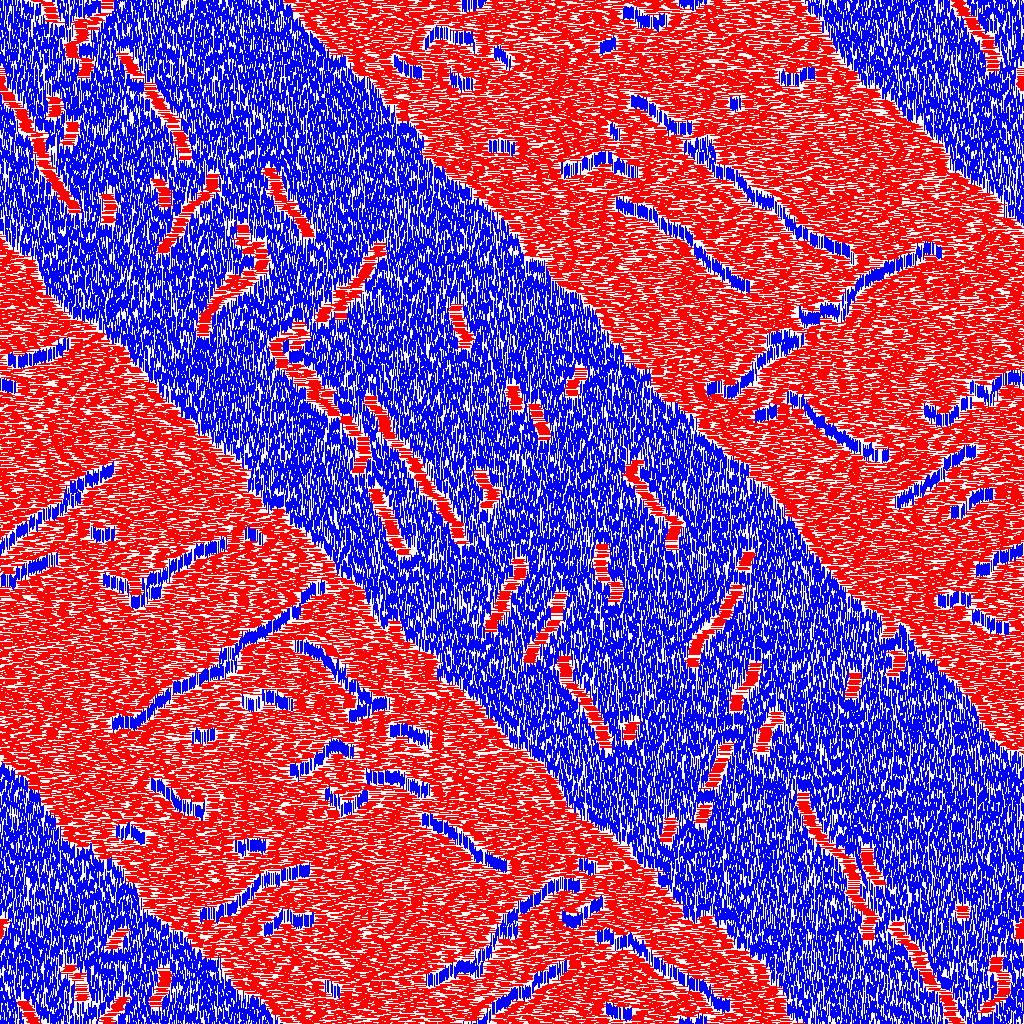}\quad
($b$)\includegraphics[width=0.425\textwidth]{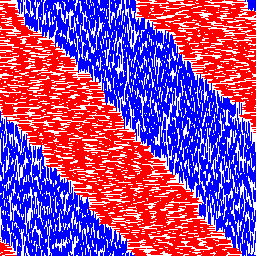}\\
($c$)\includegraphics[width=0.95\textwidth]{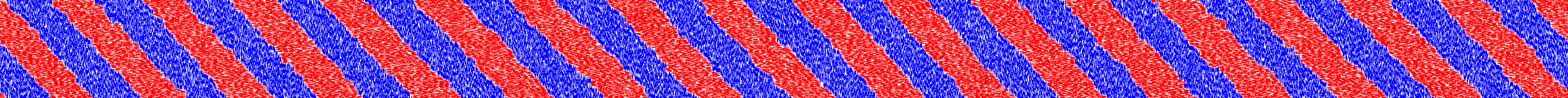}
\end{minipage}
\hfill
\begin{minipage}[b]{0.35\textwidth}
  \caption{Examples of patterns in two regions of the same area but different shapes; initial jammed state at isotropic distribution, PBCs, ``myopic'' particles, $t_\text{MC} = 10^7$, $k=12$. ($a$)~Square region of size $1024 \times 1024$. ($b$)~Fragment $256 \times 256$ of rectangular region of size $4096 \times 256$. ($c$)~Entire rectangular region of size $4096 \times 256$.}
\label{fig:patternrs}
\end{minipage}
\end{figure}

When $L_x/L_y \in \mathbb{N}$, stable stripe domains were observed at $t_\text{MC} \sim 10^7$. The width of a stripe is determined by the length of the narrower side, i.e., by the value of $L_y$. Widths of the stripes were the same when the lattices were $L_y \times L_y$ and $L_x \times L_y$. \Fref{fig:patternrs} demonstrates an example of pattern formation in an elongate rectangular region ($L_x/L_y = 16$). Final patterns in a square region with the same area is shown in~\fref{fig:patternrs}($a$) for comparison. Presence of numerous linear defects evidenced that the relaxation time exceeds  $10^7$, hence, the relaxation time is also determined by the narrower side of the region.

\begin{figure}[!htbp]
\begin{minipage}[c]{0.55\textwidth}
($a$)\includegraphics[width=0.9\textwidth]{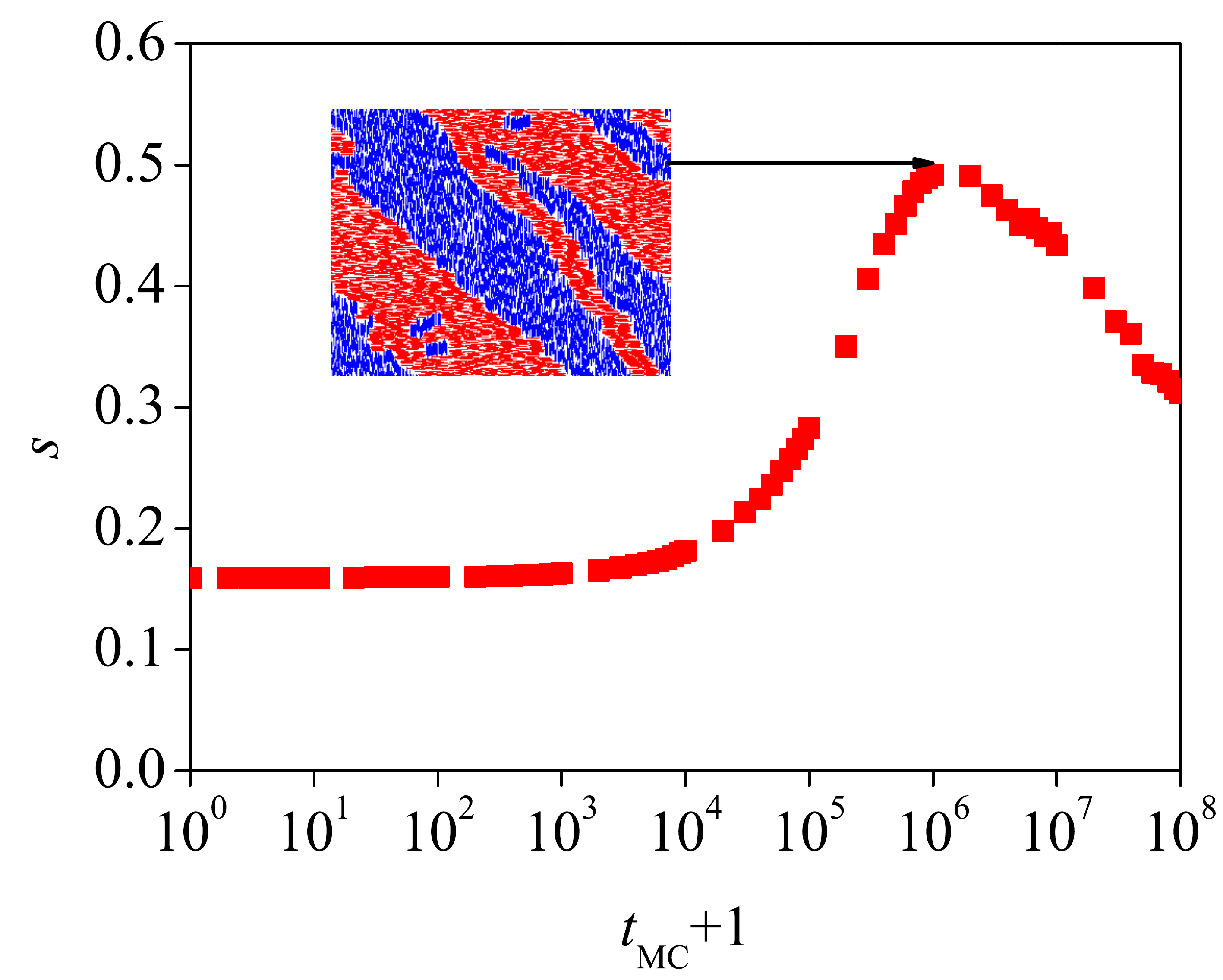}\\
($b$)\includegraphics[width=0.25\textwidth]{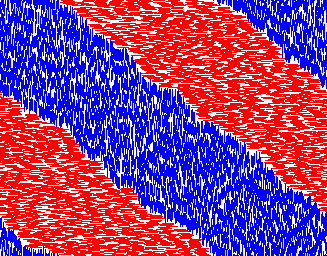}($c$)\includegraphics[width=0.25\textwidth]{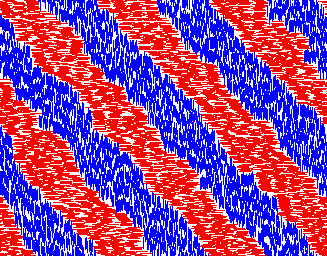}($d$)\includegraphics[width=0.25\textwidth]{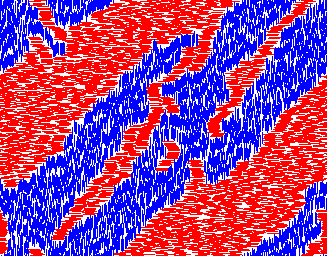}
\end{minipage}
\hfill
\begin{minipage}[c]{0.4\textwidth}
\caption{\label{fig:RG256x327s}Example of pattern evolution vs MC steps, $t_\text{MC}$, for $k = 12$. Lattice size is $327 \times 256$. 100 independent statistical runs.
($a$)~dependence of the local order parameter, $s$, calculated in windows of size $64\times 64$ sites; inset: example of a typical pattern at $t_\text{MC}=10^6$, ($b,c,d$)~examples of patterns at $t_\text{MC}=10^8$. Patterns similar to ($b$) were observed 3 times, patterns similar to ($c$) were observed 84 times, patterns similar to ($d$) were observed 13 times.}
\end{minipage}
\end{figure}

Quite different situation was observed when $L_x$ and $L_y$ are incommensurate. No steady state was observed even until $t_\text{MC} = 10^8$. By contrast,  the local order parameter was still changing at this time. Different patterns were observed at this time with different frequencies. For instance, narrow stripes were mainly observed  when $L_x/L_y = 1.27734375\dots$ (\fref{fig:RG256x327s}).

\begin{figure}[!htbp]
\begin{minipage}[c]{0.55\textwidth}
($a$)\includegraphics[width=0.9\textwidth]{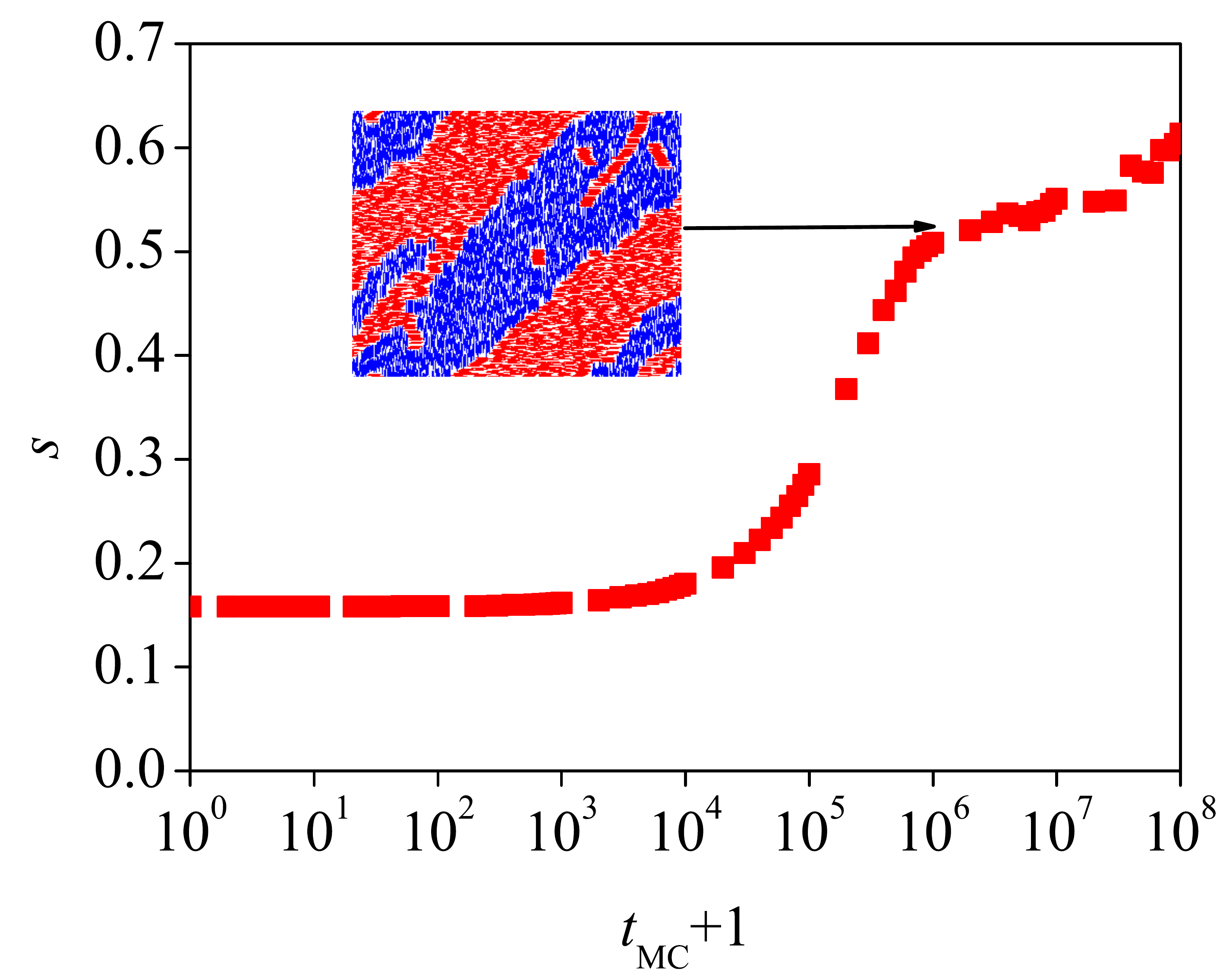}\\
($b$)\includegraphics[width=0.25\textwidth]{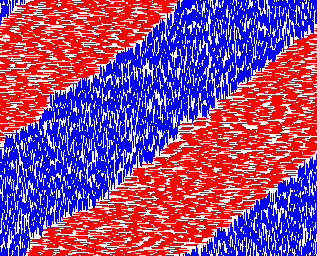}($c$)\includegraphics[width=0.25\textwidth]{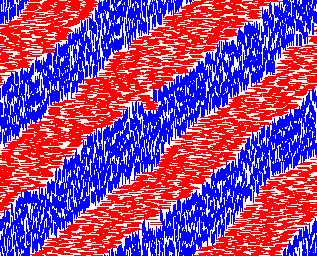}($d$)\includegraphics[width=0.25\textwidth]{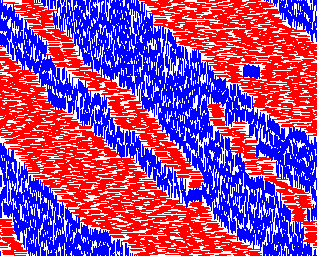}
\end{minipage}
\hfill
\begin{minipage}[c]{0.4\textwidth}
\caption{\label{fig:RG256x317s}Example of pattern evolution vs MC steps, $t_\text{MC}$, for $k = 12$. Lattice size is $317 \times 256$. 100 independent statistical runs.
($a$)~dependence of the local order parameter, $s$, calculated in windows of size $64\times 64$ sites; inset: example of a typical pattern at $t_\text{MC}=10^6$, ($b,c,d$)~examples of patterns at $t_\text{MC}=10^8$. Patterns similar to ($b$) were observed 58 times, patterns similar to ($c$) were observed 4 times, patterns similar to ($d$) were observed 38 times.}
\end{minipage}
\end{figure}

Wide stripes were mainly observed when $L_x/L_y = 1.23828125\dots$ (\fref{fig:RG256x317s}).

\begin{figure}[!htbp]
\begin{minipage}[c]{0.55\textwidth}
($a$)\includegraphics[width=0.9\textwidth]{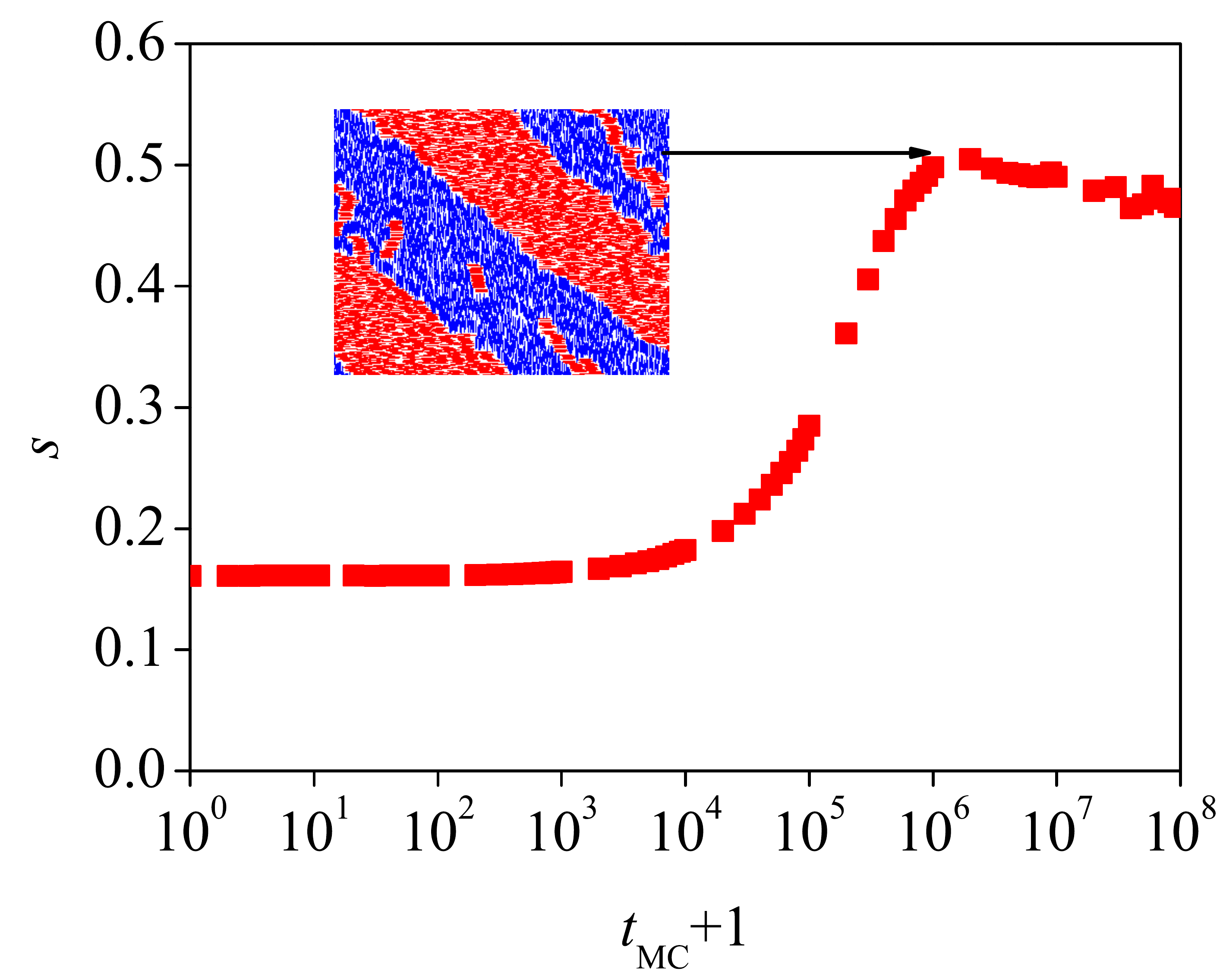}\\
($b$)\includegraphics[width=0.25\textwidth]{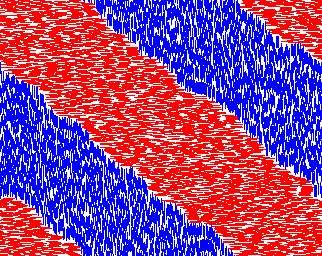}($c$)\includegraphics[width=0.25\textwidth]{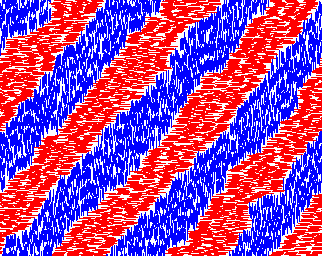}($d$)\includegraphics[width=0.25\textwidth]{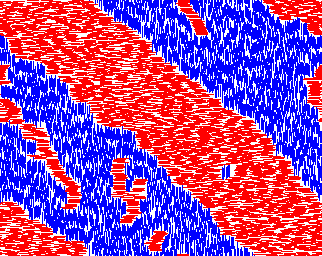}
\end{minipage}
\hfill
\begin{minipage}[c]{0.4\textwidth}
\caption{\label{fig:RG256x322s}Example of pattern evolution vs MC steps, $t_\text{MC}$, for $k = 12$. Lattice size is $322 \times 256$. 100 independent statistical runs.
($a$)~dependence of the local order parameter, $s$, calculated in windows of size $64\times 64$ sites; inset: example of a typical pattern at $t_\text{MC}=10^6$, ($b,c,d$)~examples of patterns at $t_\text{MC}=10^8$. Patterns similar to ($b$) were observed 22 times, patterns similar to ($c$) were observed 33 times, patterns similar to ($d$) were observed 45 times.}
\end{minipage}
\end{figure}

Distorted stripes were mainly observed when $L_x/L_y = 1.2578125\dots$  (\fref{fig:RG256x322s}).

We may suggest that any steady state is unreachable for systems with incommensurate aspect ratio, highly likely, such the system is permanently rebuilding its patterns.
%\begin{figure}[ht]
%\begin{minipage}[t]{0.45\textwidth}
%\includegraphics[width=\textwidth]{DCL256}
%\caption{\label{fig:DCL256}Example of dependence of the mean squared displacement, $\langle r^2 \rangle$, vs MC steps, $t_\text{MC}$, $k = 12$.}
%\end{minipage}
%\end{figure}

\section{Conclusion}\label{sec:concl}
Using Monte Carlo simulation and the lattice approach we found out that
\begin{itemize}
  \item pattern formation was observed only for particles $1 \times k$; no patterns were observed for particles with both sides greater than 1;
  \item ``blind'' and ``myopic'' particles demonstrated similar behaviour, nevertheless the final patterns are different;
  \item random walk of particles led the system in a state when the particles have more possibilities to move than in the initial jammed state;
  \item steady states were observed only for the lattices $L_x/L_y \in \mathbb{N}$, a steady state was not reached even until $t_\text{MC} = 10^8$ for incommensurable aspect ratio.
\end{itemize}

\ack The reported study was funded by RFBR according to the research project No.~18-07-00343.

\section*{References}
\bibliographystyle{iopart-num}
\bibliography{patterns}

\end{document}